\documentclass[twocolumn,aps,pra,10pt,superscriptaddress,floatfix,showkeys]{revtex4-1}
\usepackage{graphicx,subfigure}
\usepackage{amsmath,amssymb,bm}
\usepackage{mathtools}
\usepackage{multirow}
\usepackage{makecell}
\usepackage{textcomp}
\usepackage{float}
\usepackage{color}
\usepackage[normalem]{ulem}
\usepackage{dsfont}
\usepackage{mathrsfs}
\usepackage{braket}
\usepackage{transparent}
\usepackage{xcolor}
\usepackage{hhline}
\usepackage{graphicx}
\usepackage{pdfpages}
\usepackage[export]{adjustbox}
\usepackage{hyperref}
\usepackage{physics}

\makeatletter
\AtBeginDocument{\let\LS@rot\@undefined}
\makeatother

\begin{document}

\title{Gate Synthesis in Dephasing-Limited Nitrogen-Vacancy Ensembles}

\author{
Ankita~Chakravarty$^{1,2,*}$,
Arnaud~Carignan-Dugas$^{3}$,
Eva~Dupont-Ferrier$^{1,2}$,
and, Yves~Bérubé-Lauzière$^{1,4}$
\\
\medskip
\normalsize{$^{1}$Institut quantique, Université de Sherbrooke, Sherbrooke, Québec, J1K 2R1, Canada}\\
\normalsize{$^{2}$Département de physique, Université de Sherbrooke, Sherbrooke, Québec, J1K 2R1, Canada}\\
\normalsize{$^{3}$Keysight Technologies Canada, Kanata, ON, K2K 2W5, Canada}\\
\normalsize{$^{4}$Département de génie électrique et de génie informatique, Université de Sherbrooke, Sherbrooke, Québec, J1K 2R1, Canada}\\
\medskip
\small{$^{*}$Corresponding author: Ankita Chakravarty (email: Ankita.Chakravarty@USherbrooke.ca)}\\
}

\begin{abstract}

Ensemble nitrogen-vacancy (NV) centers are widely used for quantum sensing and magnetometry, where performance depends on the accurate implementation of microwave pulse sequences. In these systems, inhomogeneous broadening and short coherence times limit reliable multi-pulse control, while hardware imperfections and off-resonant excitation distort implemented rotations, making direct calibration of arbitrary gates tedious. To address this, we implement a control framework with a ZXZXZ gate decomposition which uses fixed-angle, phase-parameterized pulses, enabling universal single-qubit control from a single calibrated $
\pi/2$ primitive ($X_{\pi/2}$) along with virtual Z operations. We compare a standard pulse with a composite pulse designed for high-power operation in a regime where the Rabi frequency is comparable to the hyperfine splitting, leading to non-uniform evolution across hyperfine levels. Using pre-determined combinations of $\pi/2$ pulse sequences, we observe that standard pulses rapidly deviate from ideal behavior under repeated application, while the composite pulse suppresses this error accumulation by more than 70$\%$ for longer sequences despite its longer duration. In contrast, randomized benchmarking yields similar decay factors for both implementations, highlighting its limited sensitivity to coherent control errors. These results establish a practical route to robust control in short-coherence ensemble systems and emphasize the importance of pulse-level characterization.

\end{abstract}


\keywords{composite pulses, ensemble magnetometry, gate decomposition, microwave control, nitrogen-vacancy centers, quantum control, quantum sensing, randomized benchmarking}

\maketitle

\maketitle


\section{Introduction}
\label{sec:introduction}

Ensembles of nitrogen-vacancy (NV) centers in diamond are a widely used platform for quantum sensing and magnetometry~\cite{Degen2017,Taylor2008,Barry2020}. While ensembles provide enhanced sensitivity through increased signal collection, they also suffer from inhomogeneous broadening and short coherence times $T_2^*$~\cite{Levchenko2015,Blinder2024}. These constraints limit the duration and complexity of microwave (MW) pulse sequences used in sensing protocols. Moreover, as sequences become longer, imperfections in the pulse implementations accumulate and directly impact signal contrast and sensitivity~\cite{Wang2012}. In this regime, pulse errors do not simply lead to a reduction in contrast: imperfect refocusing can lead to faster loss of coherence~\cite{Cory2010}. As a result, control performance is governed not only by single-pulse accuracy, but by how errors accumulate under repeated pulse application.

In our system, short coherence times ($T_2^* \sim 1~\mu$s) require short-duration MW pulses operated at high power, corresponding to Rabi frequencies on the order of $3$~MHz, comparable to the $^{14}$N hyperfine splitting ($\sim 2.16$~MHz)~\cite{Steiner2010,Felton2009}. This leads to off-resonant excitation, rotation-axis distortion, and non-uniform evolution across hyperfine levels. Under these conditions, directly calibrating arbitrary gates becomes impractical, as the effective rotation depends on both the control parameters and the underlying inhomogeneity, and can vary across sequences. Techniques such as pulse shaping and optimal control~\cite{Anderson2022,Casanova2025} can, in principle, compensate for inhomogeneous broadening and control imperfections. However, these approaches typically require detailed system characterization and MW pulse waveform optimization, and may not remain robust under strong inhomogeneity and hardware constraints. 

To address this, we adopt a control framework based on a ZXZXZ gate decomposition, which uses fixed-angle, phase-parameterized MW pulses. Rather than independently calibrating the physical rotations required for different gates, we calibrate a single $\pi/2$ rotation around the $X$ axis ($X_{\pi/2}$ primitive) and construct arbitrary single-qubit operations through phase control and virtual $Z$ rotations~\cite{McKay2017,Barenco1995}. The virtual $Z$ rotations require no additional physical MW pulses, allowing universal control using a constrained set of experimentally realizable operations while incorporating systematic imperfections into a single effective primitive.

Within this framework, control performance is strongly determined by the quality of the calibrated $X_{\pi/2}$ primitive. We therefore compare a standard pulse with a composite pulse designed to implement an $X_{\pi/2}$ rotation under high-power operation in the presence of hyperfine structure. Composite pulse techniques, originally developed in nuclear magnetic resonance~\cite{Levitt1986,Vandersypen2005}, have been applied to NV ensembles to mitigate systematic control errors ~\cite{Aiello2013,Dong2021,Dong2022,Dong2021_2}. Using various combinations of $\pi/2$ pulses, we directly probe error accumulation over multiple pulse applications and observe that standard pulses rapidly deviate from ideal behavior, while the composite pulse more closely tracks ideal evolution despite its longer duration. Interestingly, randomized benchmarking (RB) yields similar decay factors for both implementations, indicating that it is largely insensitive to these coherent control errors. These results show that, in short-coherence ensemble systems, control performance must be evaluated at the level of pulse primitives and their behavior under repeated application, where accurate implementation of pulse sequences is essential.

The remainder of this paper is organized as follows. In Sec. \ref{sec:nv_background}, we introduce the experimental NV center setup and its control architecture. In Sec. \ref{sec:theoretical_background}, we present the gate synthesis framework and composite pulse construction, together with their experimental implementation. In Sec.~\ref{sec:characterization}, we discuss the control characterization results. Finally, in Sec.~\ref{sec:Discussion}, we summarize our findings and future work.

\section{Experimental Setup}
\label{sec:nv_background}

\subsection{Device Description}

The experiments are conducted using a compact, fully integrated NV ensemble magnetometry platform as described in~\cite{Chakravarty2026}. The negatively charged NV center in diamond is a spin-1 defect, consisting of a substitutional nitrogen atom adjacent to a lattice vacancy. The diamond sample (Element Six, DNV-B1) is a single-crystal diamond of dimensions $1 \times 1 \times 0.5~\mathrm{mm}^3$ with an NV concentration of approximately 300~ppb and an inhomogeneous dephasing time of $T_2^* \sim 1~\mu\mathrm{s}$.

The ground-state Hamiltonian \cite{Barry2020} of the $\text{NV}^-$ center is given by
\begin{equation}
\begin{split}
\frac{H}{h} = 
\Delta S_z^2 + \gamma_e (\mathbf{B} \cdot \mathbf{S})
\;+\;
\mathbf{S} \cdot \mathbf{A} \cdot \mathbf{I}
+ P I_z^2
- \gamma_n (\mathbf{B} \cdot \mathbf{I}),
\end{split}
\label{eq:hamiltonian_NV}
\end{equation}
where $\Delta \approx 2.87$~GHz is the zero-field splitting, $\gamma_e (\mathbf{B} \cdot \mathbf{S})$ describes the Zeeman interaction with an external magnetic field $\mathbf{B}$, $\mathbf{S} \cdot \mathbf{A} \cdot \mathbf{I}$ represents the hyperfine coupling to the host $^{14}$N nuclear spin where $\mathbf{A}$ is the nuclear hyperfine tensor, $P I_z^2$ represents the nuclear quadrupole interaction, and $\gamma_n(\mathbf{B}\cdot\mathbf{I})$ represents the nuclear Zeeman interaction. Hyperfine coupling to the host $^{14}$N nucleus results in a triplet splitting of approximately 2.16~MHz on each electronic spin transition \cite{Felton2009}. In the regime explored here, this hyperfine structure plays a significant role in determining the effective MW-driven dynamics.

\subsection{Control Architecture}

\begin{figure}[t]
  \centering
    \includegraphics[width=1\linewidth]
    {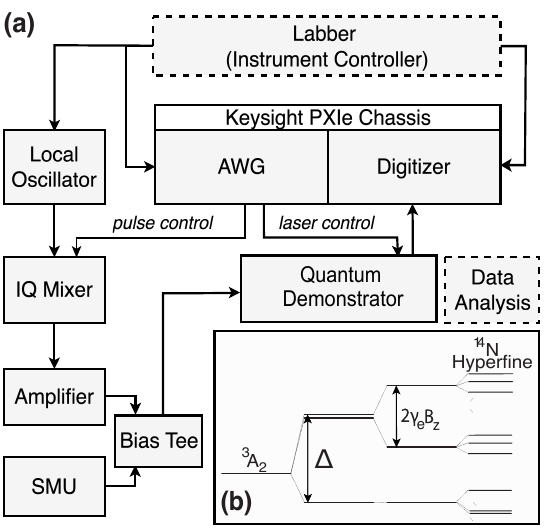}
    \caption{(a) Schematic of the experimental setup, adapted from \cite{Chakravarty2026} under the terms of the Optica Open Access Publishing Agreement. The AWG produces the MW control signals, which are upconverted using an IQ mixer driven by a local oscillator. The resulting signal is amplified and delivered to the NV ensemble inside the Quantum Demonstrator platform developed by SBQuantum. Spin-state-dependent fluorescence is collected optically and digitized using a high-speed digitizer to obtain the measured signal. (b) Energy level diagram of the ground state of the negatively charged NV center in diamond. The ground-state spin triplet ($^{3}A_{2}$) exhibits a zero-field splitting of $\Delta = 2.87$~GHz between the $m_s = 0$ and $m_s = \pm1$ states. An external bias magnetic field $B_z = 2$~mT lifts the degeneracy of the $m_s = \pm1$ levels, enabling selective MW addressing of the corresponding transitions. Optical pumping with a 532~nm green laser initializes the spin into the $m_s = 0$ state. Hyperfine coupling to the host $^{14}$N nuclear spin further splits each electronic transition into three lines separated by approximately 2.16~MHz.      
}
  \label{fig:setup}
\end{figure}

The NV ensemble is controlled using a combination of optical excitation and MW pulses. Following optical initialization, MW control of the spin-triplet ground state enables coherent spin manipulation~\cite{Doherty2013}. Each measurement step consists of two consecutive sequences, with MW on and MW off, respectively. Following this, the spin-state-dependent fluorescence is collected, and the difference between the fluorescence signals from the two sequences (MW on - MW off) is used as the measured "Signal Difference".

MW sequences are implemented using an arbitrary waveform generator (AWG; Keysight M3202A), enabling software-defined control over pulse amplitude, duration, frequency, and phase. The AWG generates in-phase ($I$) and quadrature ($Q$) components at an intermediate frequency (IF) $f_{\mathrm{IF}} = 100~\mathrm{MHz}$, combined as
\begin{equation}
V_{\mathrm{AWG}}(t) = I(t)\cos(2\pi f_{\mathrm{IF}} t) + Q(t)\sin(2\pi f_{\mathrm{IF}} t),
\end{equation}
which are upconverted using an IQ mixer (Keysight U3022A H37) driven by a local oscillator (Keysight PSG E8275D), producing a MW drive at frequency $f_{\mathrm{MW}} = f_{\mathrm{LO}} + f_{\mathrm{IF}}$. The signal is subsequently amplified and delivered to the NV ensemble inside the Quantum Demonstrator (Fig. \ref{fig:setup}(a)).

This architecture enables phase-coherent pulse sequencing, where the relative phase between successive pulses defines the rotation axis on the Bloch sphere. As a result, residual imperfections in the IQ chain—such as gain imbalance, quadrature phase error, and phase drift—map directly as systematic rotation-axis errors.

In addition, the short coherence time imposes a stringent requirement on MW control, requiring short, high-power pulses. In this system, the maximum achievable Rabi frequency is approximately 3~MHz, which is comparable to the $^{14}$N hyperfine splitting. This leads to near-resonant off-axis excitation and distortion of ideal transverse rotations. In particular, the detuning of the hyperfine transitions from the applied MW frequency causes different components of the ensemble to undergo rotations about different effective axes. Inhomogeneous broadening further introduces a distribution of detunings and hence a distribution of effective rotations, while imperfections in the MW phase control can introduce additional deviations from the intended rotation. As a result, the physically implemented operations can deviate from ideal single-axis rotations and are more accurately described by an effective tilted-axis model. MW pulse sequences used for sensing involve repeated application of such operations, making the system particularly sensitive to coherent error accumulation. This motivates the use of gate decomposition and composite pulse strategies.

\section{Gate Synthesis}\label{sec:theoretical_background}

\subsection{Decomposing single-qubit gates using phase-parameterized gate primitives}

\begin{figure}[t]
  \centering
    \includegraphics[width=1\linewidth]
    {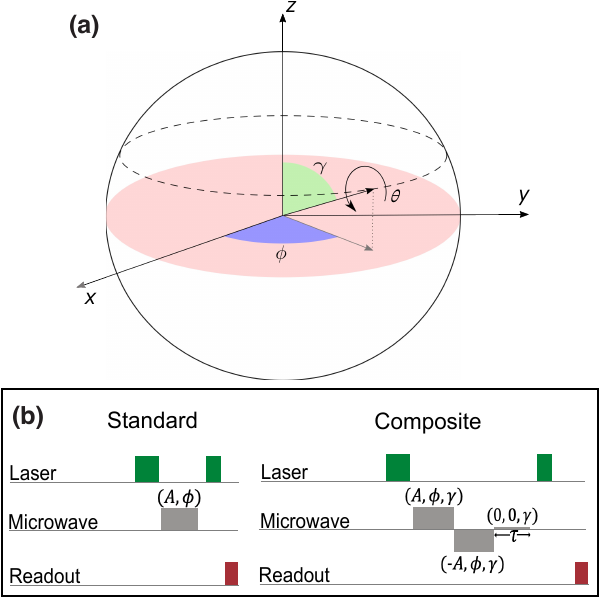}
    \caption{(a) Bloch sphere representation of the effective single-qubit rotation generated by the MW control Hamiltonian in Eq.~\eqref{eq:hamiltonian}. The phase $\phi$ defines the ideal rotation axis in the equatorial ($xy$) plane, $\gamma$ characterizes the tilt of the experimentally realized rotation axis away from this ideal axis, and $\theta$ denotes the rotation angle about the tilted experimental axis, as described by the parametrized primitive $V(\theta,\phi,\gamma)$ used throughout this work. (b) MW pulse sequences used to implement the calibrated $\pi/2$ primitive: The standard implementation consists of a single rectangular MW pulse of duration 54~ns. The composite implementation replaces this with a structured sequence of two equal-duration pulses with opposite amplitudes $(+A,-A)$ followed by a zero-amplitude wait time, resulting in a total duration of 174~ns. A global phase offset $\gamma = -2.9$~rad is applied uniformly across all segments. These primitives form the building blocks for all gate decompositions used in this work.}
  \label{fig:BS_gate}
\end{figure}

To formalize pulse-level control, we describe single-qubit operations using phase-parameterized MW pulses. For an ideal resonant drive, the effective Hamiltonian ($\hbar = 1$) generating a rotation around an axis in the equatorial plane of the Bloch sphere is given by
\begin{equation}
H_{XY} = \frac{\Omega}{2} \left( \cos\phi X + \sin\phi Y \right), 
\end{equation}
where $\Omega$ is the angular Rabi frequency and $\phi$ is the relative MW phase, which determines the rotation axis in the equatorial plane (see Fig. \ref{fig:BS_gate}(a)). For a pulse of duration $t$, the corresponding rotation angle is $\theta = \Omega t$, and the resulting unitary is
\begin{align}\label{eq:uxy}
    U(\theta, \phi) = e^{i H_{XY} t} = 
    \begin{bmatrix}
        \cos(\theta/2) & i e^{-i\phi}\sin(\theta/2) \\
        ie^{i\phi}\sin(\theta/2)  & \cos(\theta/2)
    \end{bmatrix}~.
\end{align}

In this parameterization, $X = U(\pi,0)$ and $Y = U(\pi,\pi/2)$ (up to a global phase). $Z$ rotations (Appendix \ref{appendix:pulse_decomp_theory}) are implemented virtually via frame updates, which shift the phase reference of all the subsequent $XY$ pulses ~\cite{McKay2017, Chen2023,Lidar2025}

\begin{equation}\label{eq:z}
Z(\phi) =
\begin{bmatrix}
1 & 0 \\
0 & e^{i\phi}
\end{bmatrix}.
\end{equation}
Since these virtual $Z$ operations are applied in software, they therefore introduce no additional duration or decoherence.

It can be seen that, 
\begin{align} \label{eq:U_theta_phi}
    U(\theta, \phi) = Z(\phi) U(\theta,0) Z(-\phi)~,
\end{align}
which geometrically implies that $Z$ rotates the Bloch sphere around the $z$-axis, turning the $x$-axis into a rotated direction in the equatorial plane. Using Eq. \eqref{eq:U_theta_phi} gives
\begin{align*}
    U(\theta, \phi) Z(\phi_0) = Z(\phi)U(\theta,0)Z(-\phi)Z(\phi_0).
\end{align*}
Combining the $Z$'s and regrouping results in
\begin{align*}
    U(\theta, \phi) Z(\phi_0) = Z(\phi_0)(Z(-\phi_0+\phi)U(\theta,0)Z(\phi_0 - \phi)),
\end{align*}
which gives
\begin{align*}
    U(\theta, \phi) Z(\phi_0) = Z(\phi_0)U(\theta, \phi-\phi_0).
\end{align*}
Applying this repeatedly allows phase shifts to be performed virtually such that
\begin{multline}
U(\theta_k, \phi_k) \cdots U(\theta_1, \phi_1)\, Z(\phi_0) \\
\shoveleft{= Z(\phi_0)\, U(\theta_1, \phi_k-\phi_0)\cdots U(\theta_k, \phi_1-\phi_0)}.
\end{multline}
This implies that a $Z$ rotation does not require a physical pulse. Instead, we can shift the phase reference of all the subsequent $XY$ pulses and simply update $\phi \rightarrow \phi - \phi_0$. Since no implementation at the hardware level is involved, such $Z$ rotations are referred to as virtual $Z$ gates. Unlike $XY$ rotations, which require real MW pulses, these $Z$ rotations are implemented by redefining the drive phase and are therefore effectively error-free. 
 
In the experimental platform considered here, the off-resonant $^{14}$N hyperfine transitions, together with inhomogeneous broadening and hardware imperfections that affect the realized microwave phase, can distort the effective rotation axis away from the equatorial plane. As a result, the physically implemented operation is more adequately described by a tilted-axis primitive,
\begin{multline}\label{eq:primitive}
V(\theta,\phi,\gamma) =\\
\shoveleft{
 \begin{bmatrix}
\cos(\tfrac{\theta}{2}) + i\sin(\tfrac{\theta}{2})\cos(\gamma) &
i e^{-i\phi}\sin(\tfrac{\theta}{2})\sin(\gamma) \\
i e^{i\phi}\sin(\tfrac{\theta}{2})\sin(\gamma) &
\cos(\tfrac{\theta}{2}) - i\sin(\tfrac{\theta}{2})\cos(\gamma)
\end{bmatrix}.
}
\end{multline}
The unitary $V(\theta,\phi,\gamma)$ serves as the gate primitive, with the pulse envelope determining the rotation angle $\theta$, the pulse phase relative to the rotating frame determining the azimuthal angle $\phi$, and $\gamma$ capturing the rotation-axis tilt (see Fig. \ref{fig:BS_gate}(a)).

This tilt is due to the non-uniform response of the NV ensemble to the MW drive, arising mainly from inhomogeneous broadening and the $^{14}$N hyperfine structure. In particular, near-resonant driving causes different components of the ensemble to experience different detunings and hence different effective rotation axes. Since the Rabi frequency is comparable to the hyperfine splitting, the resulting non-uniform evolution across the hyperfine levels is significant and produces a net deviation from the ideal rotation around an axis in the equatorial plane. Imperfections in the MW control chain (e.g. IQ mixer miscalibration) can further alter the phase of realized operation, with the resulting effective tilt captured by the device-dependent parameter $\gamma$. This tilted-axis operation can be obtained by considering the more general Hamiltonian
\begin{align}\label{eq:hamiltonian}
H = \frac{\Omega}{2} \left[\sin(\gamma) \Big( \cos(\phi) X + \sin(\phi) Y \Big) + \cos(\gamma) Z \right]~.
\end{align}
For $\gamma = \pi/2$, the Hamiltonian reduces to $H_{XY}$. On the Bloch sphere, the evolution $e^{iHt}$ corresponds to a rotation by an angle $\theta$ about the axis shown in Fig. \ref{fig:BS_gate}(a).

Because the available hardware constrains both the rotation angle and axis, arbitrary single-qubit gates cannot be implemented by direct calibration of individual operations. Instead, we adopt a ZXZXZ decomposition to construct all gates to effectively span SU(2) from a fixed-angle primitive using phase control and virtual $Z$ rotations. In this framework, all physical control is restricted to a calibrated $X_{\pi/2}$ primitive, while phase updates with virtual $Z$ operations provide the remaining degrees of freedom required for universality. Shorter constructions such as ZXZ~\cite{Chen2023} are not generally compatible with fixed-angle constraints in the presence of rotation-axis tilt. Appendix~\ref{appendix:pulse_decomp_theory} provides the mathematical details for the ZXZXZ decomposition, which offers a practical route to universal control using experimentally realizable operations.

\subsection{Composite Pulse}

A composite pulse implements a target rotation through multiple consecutive control segments whose combined evolution is designed to reduce the sensitivity of the operation to systematic control errors. Within the ZXZXZ framework considered here, the overall control performance depends strongly on the accuracy of the underlying $X_{\pi/2}$ primitive (Fig. \ref{fig:BS_gate}(b)). We therefore compare two implementations of the $X_{\pi/2}$ operation. The first is a standard $X_{\pi/2}$ pulse consisting of a single rectangular MW pulse with amplitude $A$, phase $\phi$, and a duration of 54~ns. This serves as a baseline for evaluating error accumulation under repeated application.

The second is a composite $X_{\pi/2}$ pulse of the form $(+A, -A, 0)$, consisting of two pulses of equal duration of 60~ns with opposite amplitudes $\pm A$, followed by a zero-amplitude waiting period $\tau = 54$~ns. The durations of the driven segments and the waiting period are calibrated such that the combined evolution suppresses the contributions from the $m=\pm1$ hyperfine levels while producing the desired net $\pi/2$ rotation for the resonant $m=0$ level. The total duration of the composite pulse is therefore 174~ns (Appendix~\ref{appendix:pulse_calibration}). A global phase offset of $\gamma = -2.9$~rad is applied to all segments and is implemented through cumulative frame updates across the sequence.

This construction reduces sensitivity to hyperfine-induced distortions by partially canceling unwanted evolution during the driven segments while allowing coherent evolution during the free precession interval. Unlike conventional composite pulses such as BB1 or CORPSE~\cite{Levitt1986,Wimperis1994,Cummins2000}, which rely on longer pulse sequences, this design is tailored to the short coherence time regime and limits the temporal overhead associated with composite control. As a result, the composite pulse improves the fidelity of the effective primitive while keeping the additional duration substantially shorter than that of conventional composite-pulse constructions, thereby limiting additional dephasing \cite{Zhang2026}.

\section{Control Characterization Results}
\label{sec:characterization}

\begin{figure*}[t]
  \centering
    \includegraphics[width=1\linewidth]
    {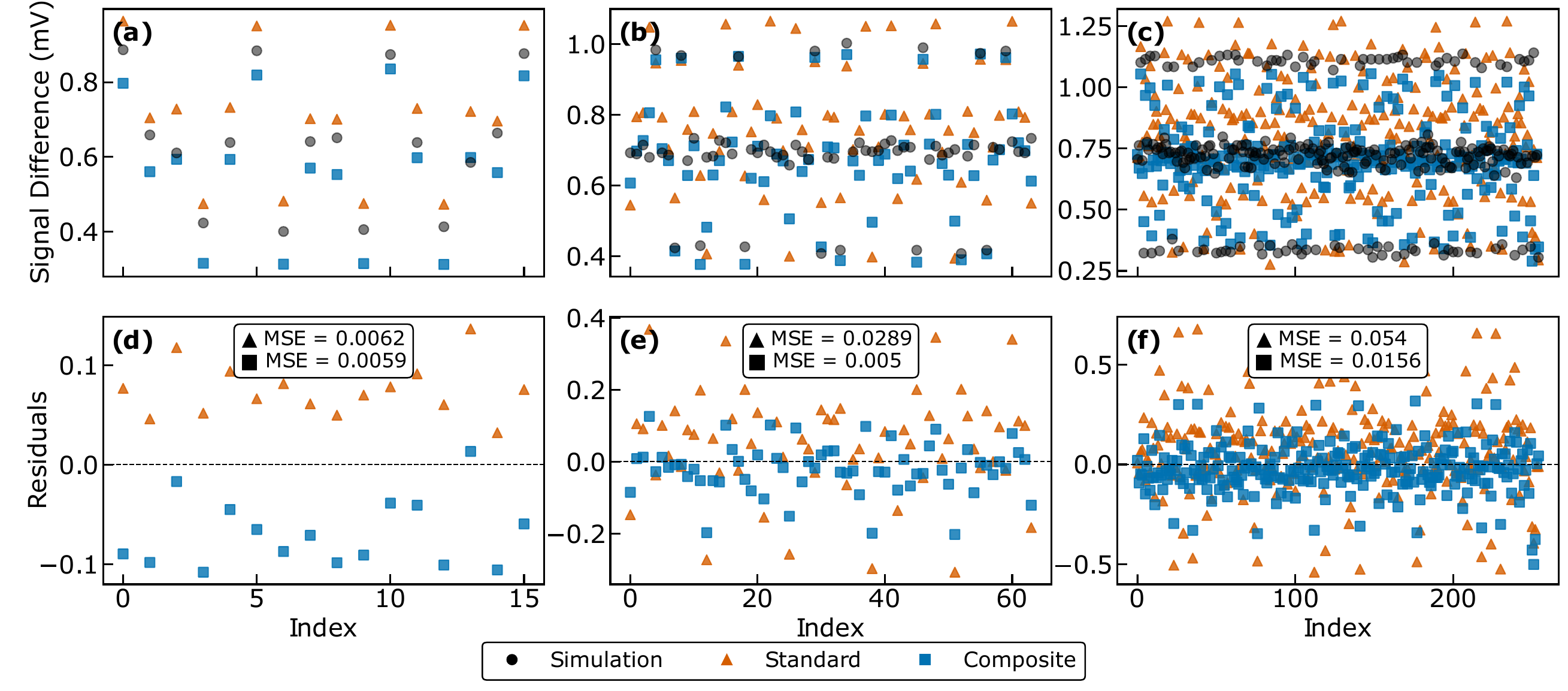}
    \caption{Comparison of structured pulse sequences implemented with standard and composite $\pi/2$ pulses, alongside ideal numerical simulations: (a) Two-gate sequences, (b) Three-gate sequences, (c) Four-gate sequences; (d)--(f) Corresponding residuals between experiment and ideal simulation, quantified using the mean squared error (MSE). The horizontal axis, "Index", denotes the iteration number within each gate sequence. Both standard and composite pulse implementations use the same ZXZXZ framework, with the calibrated $X_{\pi/2}$ primitive providing the basis for implementing arbitrary $\pi/2$ rotations through phase control. The standard and composite pulses therefore represent two physical implementations of this $\pi/2$ primitive. While both pulse types closely follow the ideal evolution for two-gate sequences, deviations become increasingly pronounced with sequence length for standard pulses. The composite pulse substantially suppresses this error growth, reducing the MSE by approximately 83\% and 71\% for three- and four-gate sequences, respectively, indicating significantly slower accumulation of coherent control errors under repeated application. }
  \label{fig:structured_sequences}
\end{figure*}

To directly probe pulse-level coherent errors, we implement structured sequences composed of repeated $X_{\pm\pi/2}$ and $Y_{\pm\pi/2}$ rotations. We constructed all possible two-, three-, and four-gate sequences to generate $4^2 = 16$, $4^3 = 64$, and $4^4 = 256$ circuits respectively. Four identical ideal $\pi/2$ gates correspond to a full $2\pi$ rotation, providing an upper bound for resolving deviations from ideal behavior.  These sequences deliberately amplify systematic calibration errors: in particular, repeated rotations about the same axis are highly sensitive to coherent over-rotation and phase misalignment. Both standard and composite pulses are used to realize the same set of $X_{\pi/2}$ rotations in the ZXZXZ decomposition, thus ensuring that differences arise solely from the underlying pulse implementation.

Figure~\ref{fig:structured_sequences}(a)–(c) compares experimental measurements with ideal numerical simulations (Appendix~\ref{appendix:trueq_simulation}). Sequences constructed from standard $\pi/2$ pulses exhibit rapid deviation from ideal evolution. This is quantified in the residual plots in Fig.~\ref{fig:structured_sequences}(d)–(f) using the mean squared error (MSE) relative to the ideal signal. For standard pulses, the MSE increases from $0.0062$ for two-gate sequences to $0.0289$ and $0.054$ for three- and four-gate sequences, respectively. In contrast, the composite pulse yields substantially smaller values of $0.0059$, $0.0050$, and $0.0156$. While both implementations perform similarly for two-gate sequences, the composite pulse reduces the MSE by approximately $83\%$ and $71\%$ for three- and four-gate sequences, respectively, indicating significantly slower coherent error accumulation under repeated application.

\begin{figure*}[t]
  \centering
    \includegraphics[width=1\linewidth]
    {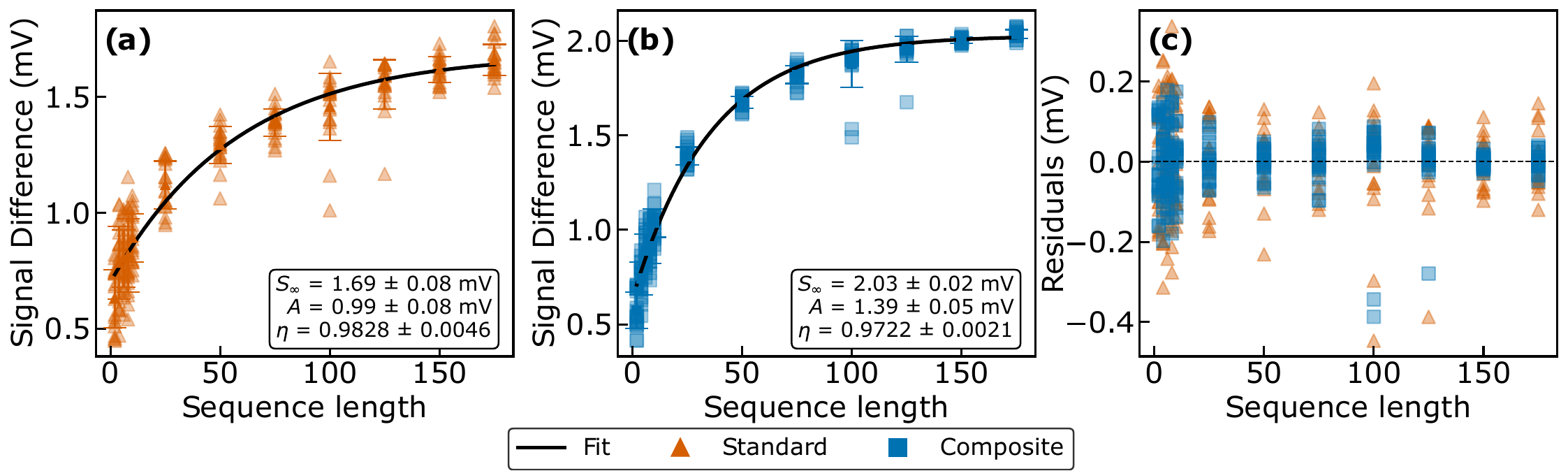}
  \caption{Randomized benchmarking (RB) measurements for sequence lengths $n = [2,4,6,8,10,25,50,75,100,125,150,175]$ Clifford gates. (a) Standard $\pi/2$ pulse implementation (54~ns per pulse). (b) Composite $\pi/2$ pulse implementation (174~ns per pulse). (c) Deviation of the measured signal from the mean across randomized circuits for each sequence length, illustrating reduced fluctuations for the composite pulse. In contrast to structured-sequence experiments, RB probes error accumulation through random Clifford gate averaging and is primarily sensitive to incoherent processes such as dephasing. Both implementations show similar overall decay trends, despite the composite pulse having a substantially longer physical duration, consistent with partial cancellation of coherent control errors within randomized sequences.}
  \label{fig:RB_normal_composite}
\end{figure*}

This difference is not apparent in randomized benchmarking (RB), a widely used protocol for characterizing gate performance~\cite{Knill2008,Feng2016}, which probes error accumulation in a fundamentally different manner. Clifford gates are implemented using the ZXZXZ decomposition described in Sec.~\ref{sec:theoretical_background}, where each Clifford gate consists of two physical $\pi/2$ pulses interleaved with virtual $Z$ operations. An RB sequence of length $n$ therefore contains $2(n+1)$ physical pulses, including the final inversion gate. Figures~\ref{fig:RB_normal_composite}(a) and (b) show the RB measurements for the standard and composite pulse implementations. Unlike conventional RB, where the measured quantity is interpreted as the survival probability, here we record the fluorescence contrast defined by the signal difference. To model this RB behavior, we fit the data using the following equation
\begin{equation}\label{eq:RB_model}
    S(n) = S_{\infty} - A \eta^n,
\end{equation}
where $S(n)$ is the measured signal difference after $n$ Clifford gates, $S_{\infty}$ is the asymptotic signal value reached for long sequence lengths, $A$ is the initial contrast amplitude, and $\eta$ is the effective depolarizing parameter.

For a purely dephasing process with coherence time $T_2$, Clifford averaging maps the decay onto an effective depolarizing parameter
\begin{equation}\label{eq:eta_t2}
    \eta_{T_2} = \frac{2}{3}e^{-t_c/T_2}+\frac{1}{3},
\end{equation}
where $t_c$ is the duration of a single Clifford gate. Since each Clifford gate is implemented using two calibrated $\pi/2$ pulses through the ZXZXZ decomposition, the effective Clifford gate durations are $t_c^{\mathrm{standard}} = 108~\mathrm{ns}$ and $t_c^{\mathrm{composite}} = 348~\mathrm{ns}$ for the standard and composite implementations, respectively.

Fitting the RB data yields $\eta^{\mathrm{standard}} \approx 0.98$ and $\eta^{\mathrm{composite}} \approx 0.97$. To estimate the dephasing-limited contribution to the observed RB decay, the extracted values of $\eta$ are compared to the dephasing-limited prediction of Eq.~\eqref{eq:eta_t2} using the experimentally measured spin-echo coherence time ($T_2 \sim 11~\mu$s). The use of $T_2$ rather than $T_2^*$ is motivated by the fact that randomized Clifford gate sequences partially average quasi-static detuning errors through repeated basis rotations, making the effective decay during RB more closely related to spin-echo-limited coherence than to free-induction decay. Using Eq.~\eqref{eq:eta_t2}, we obtain $\eta^{\mathrm{standard}}_{T_2} \approx 0.99$ and $\eta^{\mathrm{composite}}_{T_2} \approx 0.98$. Comparing these values to the experimentally observed decay, the fraction of the observed decay associated with dephasing is then estimated as
\begin{align}
f_{T_2} = \frac{1-\eta_{T_2}}{1-\eta}.
\end{align}
This gives $f_{T_2}^{\mathrm{standard}} \approx 0.50$ and $f_{T_2}^{\mathrm{composite}} \approx 0.67$, indicating that dephasing accounts for approximately $50\%$ and $67\%$ of the observed RB decay for the standard and composite implementations, respectively. The larger dephasing contribution for the composite pulse is expected given its approximately threefold longer gate duration, reflecting a trade-off between additional dephasing and improved control. At the same time, the relatively modest difference in the observed $\eta$ values suggests that the reduction in coherent pulse-level errors demonstrated by the composite implementation partially compensates for the increased exposure to decoherence resulting from its longer duration. This interpretation is consistent with the structured-sequence measurements, where the composite pulse exhibits substantially reduced coherent error accumulation under repeated application.

This distinction is further illustrated in Fig.~\ref{fig:RB_normal_composite}(c), which shows the deviation from the mean across randomized circuits for each sequence length. The composite pulse exhibits reduced fluctuations, indicating improved phase stability and greater consistency across randomized sequences.

This highlights a known limitation of RB (Appendix \ref{appendix:rb_limitations}): coherent errors are randomized by Clifford averaging~\cite{Magesan2012,Wallman2018}. As a result, RB interpretation is not trivial \cite{Proctor2017}. In contrast, structured pulse sequences preserve and amplify coherent error accumulation, making them a more sensitive probe of pulse-level control accuracy in short-coherence ensemble systems.

\section{Conclusion}\label{sec:Discussion}
This work investigates MW control in an NV ensemble regime where short coherence times, inhomogeneous broadening, and Rabi frequencies comparable to the hyperfine splitting distort ideal single-qubit rotations. Under these conditions, coherent pulse-level errors accumulate rapidly under repeated application, making direct calibration of arbitrary gates impractical.

To address this, we implemented a control framework based on ZXZXZ gate decomposition using a single calibrated, phase-parameterized primitive $V(\theta,\phi,\gamma)$ together with virtual $Z$ rotations. This reduces arbitrary single-qubit control to repeated application of a fixed experimentally realizable pulse, enabling universal control without gate-by-gate calibration.

Structured-sequence experiments show that standard $\pi/2$ pulses rapidly deviate from ideal evolution as sequence length increases, whereas the composite pulse substantially suppresses this growth despite its 3x longer duration, reducing the mean-squared deviation by more than $70\%$ for sequences longer than two gates. These results indicate that coherent pulse-level imperfections dominate sequence behavior in this regime, even when individual pulses appear well calibrated.

In contrast, RB yields similar decay factors for both pulse implementations despite the composite pulse being more than three times longer in duration. This reflects a known limitation of standard RB: coherent control errors are averaged by Clifford randomization and cannot be separated from incoherent errors. Consequently, pulse implementations that behave very differently under structured evolution may appear similar under RB.

Together, these results show that control performance in short-coherence ensemble systems is governed not only by decoherence times, but also by how accurately physically realizable pulse primitives reproduce idealized gate operations under repeated application, which cannot be fully assessed through RB decay parameters alone. Structured pulse sequences, therefore, provide a more sensitive probe of pulse-level control accuracy than RB alone. This distinction is particularly relevant for NV-based sensing protocols~\cite{Taylor2008,deLange2010,Pham2012}, where repeated MW operations are used in the measurement sequence, and coherent pulse imperfections directly impact signal contrast and hence the sensitivity.

This motivates the development of pulse-characterization techniques capable of separating coherent and incoherent error channels, such as cycle benchmarking~\cite{Erhard2019} and purity benchmarking~\cite{Wallman2015}. Extending the present framework toward optimized composite pulses and optimal-control-based pulse engineering \cite{Rembold2020} could further improve robustness for multi-pulse sensing protocols in NV ensembles.


\appendix

\section{ZXZ and ZXZXZ gate decompositions}\label{appendix:pulse_decomp_theory}

Consider the following parameterization of all single-qubit gate representations
\begin{align}
    G(\theta,\phi,\kappa, \lambda) = e^{i\lambda}\begin{bmatrix}
        \cos(\theta/2) & ie^{-i\phi}e^{i \kappa}\sin(\theta/2)  \\
        ie^{i \phi}\sin(\theta/2)   &  e^{i \kappa} \cos(\theta/2)
    \end{bmatrix}~.
\end{align}
From the definition of the XY pulse (Eq. \eqref{eq:uxy}) and the phase gate (Eq. \eqref{eq:z}), any single-qubit gate can be expressed as
\begin{align}
    G(\theta,\phi,\kappa, 0 ) & = U(\theta,\phi) Z(\kappa)  \notag\\
     & = Z(\phi) U(\theta, 0) Z(\kappa-\phi) \notag \\
     & = e^{i\phi Z}e^{i\frac{\theta}{2} X} e^{i(\kappa-\phi)Z}~.
\end{align}
In the first line, it is seen that any gate is decomposable as an XY pulse preceded by a phase shift. The last line was added because this decomposition is commonly referred to as the ZXZ decomposition. Notice that the same decomposition is generally not possible for the primitive $V(\theta, \phi, \gamma)$ defined in Eq. \eqref{eq:primitive}. This can be easily verified by considering the $(1,1)$ matrix element of $V$, for which $|V_{11}|^2 = 1-\sin^2(\gamma) \sin^2(\theta/2)$, which does not span the range $[0,1]$ when $\gamma \neq \pi/2$. Physically this implies that an off-resonance drive creates an axis tilt and this creates a restricted primitive gate set. So in such a case, virtual Z updates or additional control is required to regain universality.

It is often not desired to tune the entire family of gates parameterized as $U(\theta,\phi)$ for all $\theta$. A decomposition is therefore sought that uses only a fixed-angle primitive while retaining full single-qubit controllability, and so another fixed-angle decomposition was developed, namely the ZXZXZ decomposition. To get to that decomposition, consider two consecutive XY pulses

\begin{multline}
\label{eq:unitary_product}
U(\theta_2,\phi_2)U(\theta_1,\phi_1) \\
\shoveleft{
 = \begin{bmatrix}
\cos(\tfrac{\theta_2}{2}) & i e^{-i\phi_2}\sin(\tfrac{\theta_2}{2}) \\
i e^{i\phi_2}\sin(\tfrac{\theta_2}{2}) & \cos(\tfrac{\theta_2}{2})
\end{bmatrix}} \notag \\
\shoveleft{~~~~~\times
\begin{bmatrix}
\cos(\tfrac{\theta_1}{2}) & i e^{-i\phi_1}\sin(\tfrac{\theta_1}{2}) \\
i e^{i\phi_1}\sin(\tfrac{\theta_1}{2}) & \cos(\tfrac{\theta_1}{2})
\end{bmatrix}}
 \\
\shoveleft{
 = \begin{bmatrix}
\cos(\tfrac{\theta_1}{2})\cos(\tfrac{\theta_2}{2})
&
i e^{-i\phi_1}\sin(\tfrac{\theta_1}{2})\cos(\tfrac{\theta_2}{2})
\\
i e^{i\phi_1}\sin(\tfrac{\theta_1}{2})\cos(\tfrac{\theta_2}{2})
&
\cos(\tfrac{\theta_1}{2})\cos(\tfrac{\theta_2}{2})
\end{bmatrix}
+} \notag \\
\shoveleft{
\begin{bmatrix}
- e^{i(\phi_1-\phi_2)}\sin(\tfrac{\theta_1}{2})\sin(\tfrac{\theta_2}{2})
&
i e^{-i\phi_2}\sin(\tfrac{\theta_2}{2})\cos(\tfrac{\theta_1}{2})
\\
i e^{i\phi_2}\sin(\tfrac{\theta_2}{2})\cos(\tfrac{\theta_1}{2})
&
- e^{i(\phi_2-\phi_1)}\sin(\tfrac{\theta_1}{2})\sin(\tfrac
{\theta_2}{2})
\end{bmatrix}
}.
\end{multline}

Let's first look at the top left element which is composed of two terms: $\cos(\theta_1/2)\cos(\theta_2/2)$ and $-e^{i(\phi_1-\phi_2)}\sin(\theta_1/2)\sin(\theta_2/2)$. If we want to generate any single-qubit gate by only changing the phases $\phi_i$, we need both terms to be of length $1/2$. This way, the magnitude of the element can be parameterized by adjusting the phase $e^{i(\phi_1-\phi_2)}$. To achieve this condition, the only choice is to pick $\theta_i = \pi/2$. This simplifies the above matrix to

\begin{multline}
U(\tfrac{\pi}{2}, \phi_2)U(\tfrac{\pi}{2}, \phi_1) =
 \begin{bmatrix}
\tfrac{1 - e^{i(\phi_1-\phi_2)}}{2} 
& i \tfrac{e^{-i\phi_1} + e^{-i\phi_2}}{2} \\
i \tfrac{e^{i\phi_1} + e^{i\phi_2}}{2}  
& \tfrac{1 - e^{i(\phi_2-\phi_1)}}{2}
\end{bmatrix}
\end{multline}
which is equivalent, up to a global phase, to
\begin{multline}
\shoveleft{
\begin{bmatrix}
\sin\!\left(\tfrac{\phi_1-\phi_2}{2}\right)
&
- e^{-i\phi_1} \cos\!\left(\tfrac{\phi_1-\phi_2}{2}\right) \\
- e^{i\phi_2} \cos\!\left(\tfrac{\phi_1-\phi_2}{2}\right)
&
- e^{i(\phi_2-\phi_1)} \sin\!\left(\tfrac{\phi_1-\phi_2}{2}\right)
\end{bmatrix}
}.
\end{multline}

Hence, two $\pi/2$ pulses generate a family that already looks like a general SU(2) block. As such, if we consider adding surrounding Z phases to give full controllability, we get (up to a global phase)
\begin{align}
    &Z(-\frac{\pi}{2})U(\frac{\pi}{2}, \phi_2+\pi)U(\frac{\pi}{2}, \phi_1)Z(\phi_1-\phi_2+\kappa+\frac{\pi}{2})  \notag\\
    & = 
    \begin{bmatrix}
          \cos(\frac{\phi_1-\phi_2}{2})& ie^{-i\phi_2} e^{i\kappa} \sin(\frac{\phi_1-\phi_2}{2}) \\
         ie^{i\phi_2} \sin(\frac{\phi_1-\phi_2}{2})  &  e^{i \kappa } \cos(\frac{\phi_1-\phi_2}{2})
    \end{bmatrix}~,
\end{align}
which is a parameterization for all single-qubit gates. In terms of $G$, we have

\begin{multline}
\label{eq:G_decomposition}
G(\theta,\phi,\kappa, \lambda)\propto \\
\shoveleft{
 Z(\phi+\frac{\pi}{2}) \, U(\frac{\pi}{2}, 0) \, Z(\theta-\pi) \, U(\frac{\pi}{2}, 0) \, Z(\phi+\kappa+\frac{\pi}{2})~,
}
\end{multline}
which is the ZXZXZ decomposition. 

It turns out that as long as $\gamma \geq \pi/4$, we can also obtain a ZXZXZ decomposition with the $V$ primitive from Eq. \eqref{eq:primitive}. The main difference is that instead of using the angle $\theta = \pi/2$, we will use
$\theta_p \in [0,\pi]$ such that $\theta_p(\gamma) =  2 \arcsin[1/(\sqrt{2} \sin(\gamma))]$. This comes from imposing the same balancing conditions as before, where we want the magnitude of the diagonal element to be $1/\sqrt{2}$. So setting $|V_{11}|^2 = 1/2$, we get: $1-\sin^2(\gamma) \sin^2(\theta_p/2) = 1/2$. This gives,
\begin{align*}
    \sin({\theta_p/2}) = \frac{1}{\sqrt{2}\sin(\gamma)}
\end{align*}
For this to be $\leq$ 1, $\sin(\gamma) \geq 1/\sqrt{2}$ which gives $\gamma \geq \pi/4$.

Let's also define the function $\alpha(\gamma) := \arcsin[\cos(\gamma)/\sin(\gamma)]$. By applying two sequential $V$ pulses, 
we get
\begin{align}
    &V\left(\theta_p(\gamma), \phi_2, \gamma\right)V\left(\theta_p(\gamma), \phi_1,\gamma\right) 
     \notag \\
    & =  {\begin{bmatrix}
          \frac{e^{i \alpha(\gamma)}}{\sqrt{2}}&  \frac{ie^{-i\phi_2}}{\sqrt{2}} \\
         \frac{ie^{i\phi_2}}{\sqrt{2}} &  \frac{e^{-i \alpha(\gamma)}}{\sqrt{2}}
    \end{bmatrix}
    \begin{bmatrix}
          \frac{e^{i \alpha(\gamma)}}{\sqrt{2}}&  \frac{ie^{-i\phi_1}}{\sqrt{2}} \\
         \frac{ie^{i\phi_1}}{\sqrt{2}} &  \frac{e^{-i \alpha(\gamma)}}{\sqrt{2}}
    \end{bmatrix}} \notag \\
    &=\begin{bmatrix}
          \frac{e^{2i \alpha(\gamma)}-e^{i(\phi_1-\phi_2)}}{2}&  i\frac{e^{i(\alpha(\gamma)-\phi_1)}+e^{-i(\alpha(\gamma)+\phi_2)}}{2} \\
         i\frac{e^{-i(\alpha(\gamma)-\phi_1)}+e^{i(\alpha(\gamma)+\phi_2)}}{2}  &  \frac{e^{-2i \alpha(\gamma)}-e^{-i(\phi_1-\phi_2)}}{2} 
    \end{bmatrix} \notag \\
    &=i\left[\begin{matrix}
           -e^{i \left(\frac{\phi_1-\phi_2}{2}+\alpha(\gamma)\right)}\sin\left(\frac{\phi_1-\phi_2}{2}-\alpha(\gamma)\right)  \\
         e^{i\frac{\phi_2+\phi_1}{2}} \cos\left(\frac{\phi_1-\phi_2}{2}-\alpha(\gamma)\right)  
    \end{matrix} \right. \notag \\
&~~~~~~~~~~~~~~~~~~~~~~~~~~\left.\begin{matrix}
             e^{-i\frac{\phi_2+\phi_1}{2}} \cos\left(\frac{\phi_1-\phi_2}{2}-\alpha(\gamma)\right)  \\
          e^{-i \left(\frac{\phi_1-\phi_2}{2}+\alpha(\gamma)\right)}\sin\left(\frac{\phi_1-\phi_2}{2}-\alpha(\gamma)\right)
    \end{matrix} \right].
\end{align}
As such, if we consider
\begin{align}
    & Z(\alpha(\gamma)-\frac{\pi}{2}) V\left(\theta_p(\gamma), \phi+\pi, \gamma\right)\notag \\
    & \times V\left(\theta_p(\gamma), \phi+2\alpha(\gamma)+\theta,\gamma\right) Z(\theta+3\alpha(\gamma)+\kappa-\frac{\pi}{2})
     \notag \\
    & =  -e^{i\theta}
    \begin{bmatrix}
            \cos\left(\frac{\theta}{2}\right)
            &  
            ie^{-i\phi}e^{i \kappa}\sin\left(\frac{\theta}{2}\right)
            \\
         ie^{i\phi} \sin\left(\frac{\theta}{2}\right)  
         & e^{i \kappa}\cos\left(\frac{\theta}{2}\right)
    \end{bmatrix},
\end{align}
where $\theta_p(\gamma)$ is the fixed primitive rotation angle determined by the hardware, and $\theta$ is a constant that specifies the target rotation angle of the synthesized single-qubit gate, we get a "ZHZHZ" decomposition for single-qubit gates. Here, $H$ denotes a fixed rotation generated by the primitive $V(\theta_p(\gamma),\phi,\gamma)$ (i.e., a rotation about the tilted control axis).


\section{Pulse Calibration}
\label{appendix:pulse_calibration}

Calibrating the IQ mixer is important for minimizing distortions in the applied MW pulses. However, residual imperfections from the full signal chain (including cabling and amplification) lead to deviations that cannot be fully captured by IQ calibration parameters such as the IQ skew and ratio. To account for these effects, the composite $\pi/2$ pulse parameters are determined experimentally through systematic parameter sweeps that directly optimize the measured fluorescence signal. This approach ensures that the calibrated pulse remains robust when applied repeatedly in succession, which is critical in multi-pulse sequences where small phase and duration errors accumulate.

The composite $\pi/2$ pulse used in this work consists of three segments: two MW pulses of equal duration $t$ and equal magnitude of the amplitude $A$ (set by the maximum available Rabi frequency) but with opposite polarities (i.e., a $\pi$ phase shift between them), followed by a free evolution interval of duration $\tau$ with zero drive amplitude, and a global phase offset $\gamma$ is applied to the entire composite pulse. In practice, this global phase serves as a correction parameter that compensates for systematic phase shifts introduced by the experimental setup, while additional gate-dependent phases are applied on top of this offset.

\begin{figure}[t]
    \centering
    \includegraphics[width=1.0\linewidth]{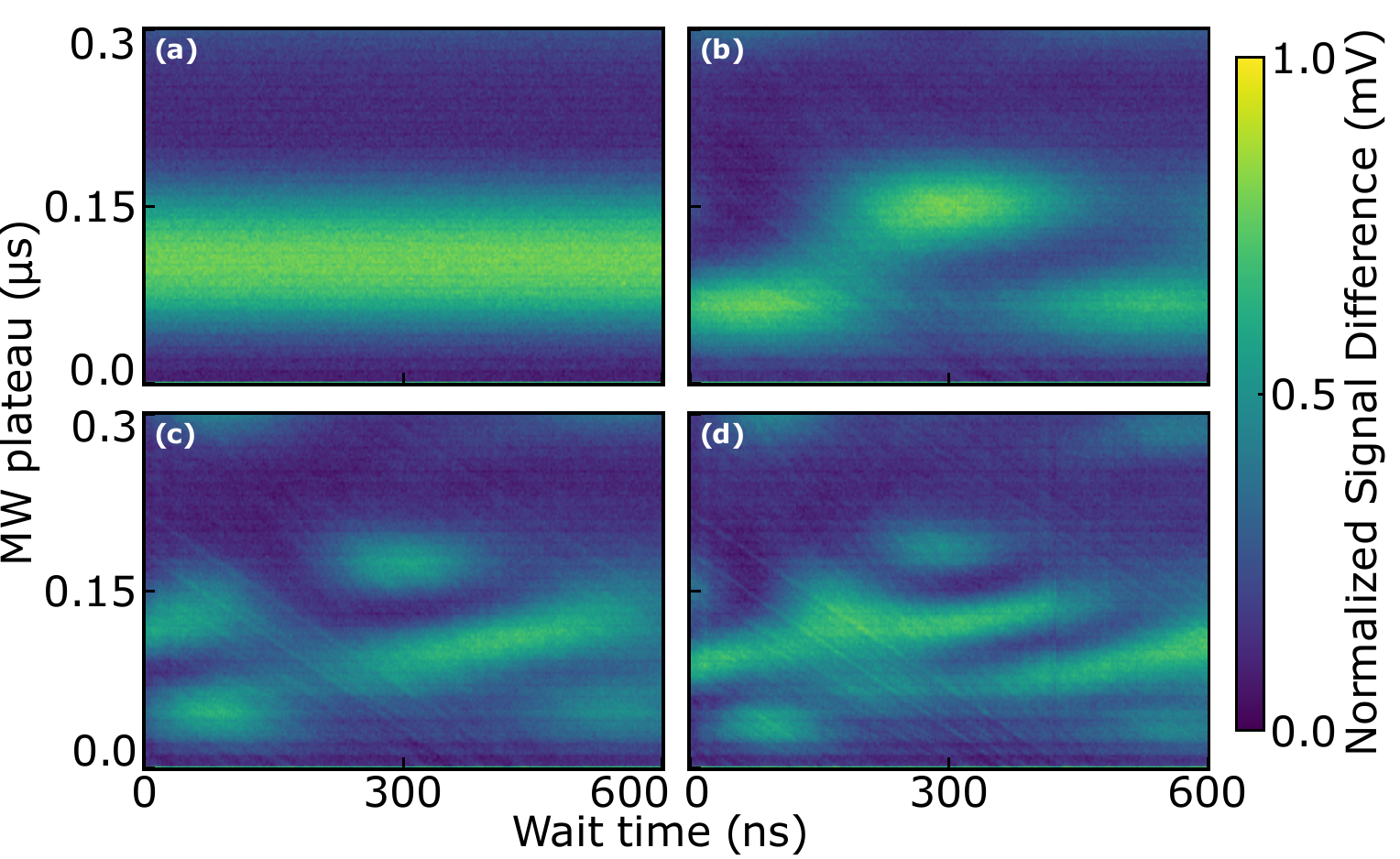}
\caption{2D calibration maps of the normalized signal difference as a function of composite pulse segment duration $t$ and wait time $\tau$, shown for sequences of (a) one, (b) two, (c) three, and (d) four repeated composite $\pi/2$ pulses.}
\label{fig:calibration_plateau_wt}
\end{figure}

The three parameters $(t, \gamma, \tau)$ are calibrated experimentally. Initial calibration of the composite $\pi/2$ pulse is performed using two-dimensional sweeps over the pulse segment duration $t$ (0--300~ns, step size 2~ns, set by the AWG resolution) and the wait time $\tau$ (0--600~ns, step size 2~ns). These sweeps are carried out for sequences consisting of one to four repeated composite pulses (Fig.~\ref{fig:calibration_plateau_wt}) in order to identify a stable operating regime under repeated application. The number of repetitions is restricted to four since four consecutive $\pi/2$ rotations correspond to a full $2\pi$ rotation on the Bloch sphere, which is sufficient to probe error accumulation over one full cycle. 

From these coarse sweeps, a region of parameter space is identified where the signal behaves as expected based on the number of $\pi/2$ pulses applied. For one pulse in Fig. \ref{fig:calibration_plateau_wt}(a), ideal operation corresponds to a normalized signal value of 0.5 (corresponding to the equator). For two pulses in Fig.~\ref{fig:calibration_plateau_wt}(b), the ideal normalized signal value is 1 (corresponding to the excited state). For three pulses in Fig.~\ref{fig:calibration_plateau_wt}(c), the ideal normalized signal value is again 0.5 (corresponding to the equator) and finally back to 0 (corresponding to the starting ground state) with the forth pulse in  Fig.~\ref{fig:calibration_plateau_wt}(d). Based on this region, we fix the MW segment duration at $t=54$~ns for the first two segments. We then perform a second set of 2D sweeps over the global phase $\gamma$ (from $-\pi$ to $\pi$ with a step size of $\pi/100$) and wait time $\tau$ (0--600~ns with a step size of 2~ns) to further refine the parameter space region. At this stage, the goal is not to identify a single optimal point, but rather to locate a stable region that can be fine-tuned based on sequence-level performance.

\begin{figure}[t]
    \centering
    \includegraphics[width=1.0\linewidth]{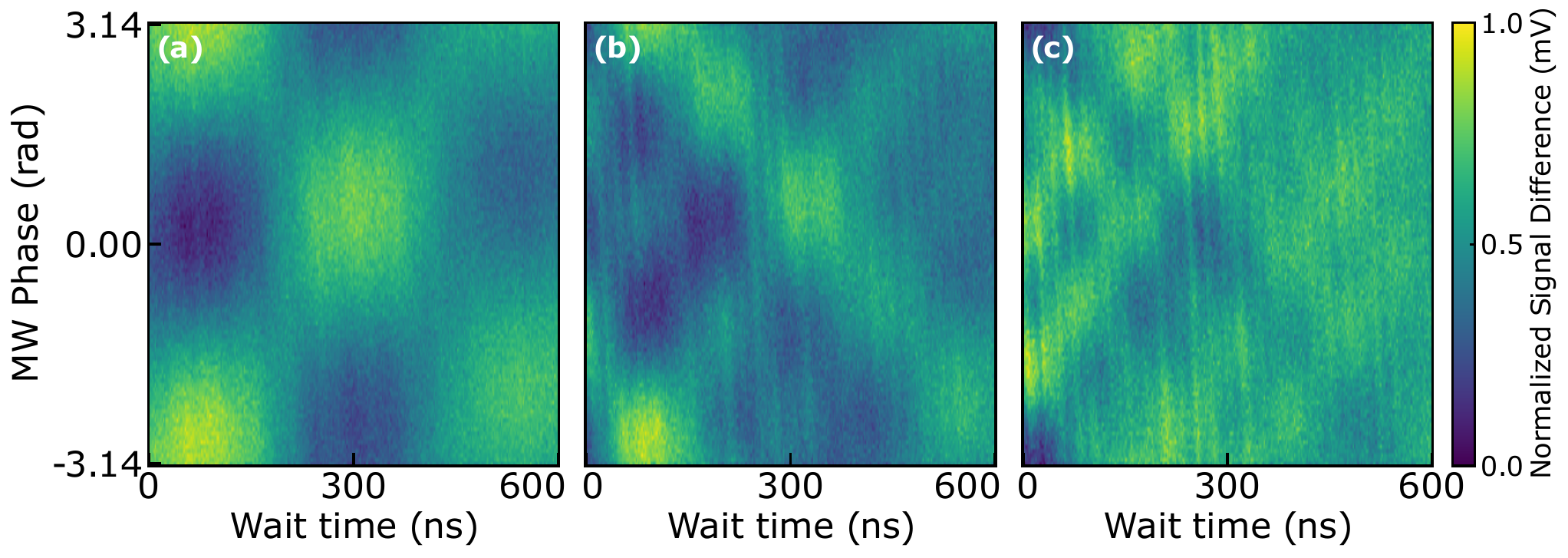}
\caption{2D calibration maps of the normalized signal difference as a function of global phase $\gamma$ and wait time $\tau$, shown for sequences of (a) two, (b) three, and (c) four repeated composite $\pi/2$ pulses.}
\label{fig:calibration_phase_wt}
\end{figure}

Following this, the parameters are fine-tuned using two- and three-gate pulse sequences by systematically sweeping over the identified parameter region and comparing the resulting experimental signals with ideal simulations of the corresponding sequences using standard $\pi/2$ pulses.
The goal is to identify parameters that give the minimum mean squared error (MSE), so that the composite pulse remains stable under repeated application, rather than parameters that optimize the pulse fidelity. From this analysis, the final composite pulse parameters are determined to be: a MW pulse duration of $t = 60$~ns for the first two segments, a wait time of $\tau = 54$~ns, and a global phase offset of $\gamma = -2.9$ rad.

All parameters are fixed after calibration and used consistently across all experiments. The resulting composite pulse provides a single calibrated implementation of the $X_{\pi/2}$ rotation, with different equatorial rotation axes and directions realized through phase control. This ensures that comparisons between different pulse implementations reflect differences in control performance rather than changes in the experimental setup.


\section{Simulation Framework}
\label{appendix:trueq_simulation}
Numerical simulations were performed using the TrueQ framework~\cite{Beale2020} to model the evolution of the structured pulse sequences used in this work. The simulations are constructed at the gate level and include experimentally relevant parameters such as the inhomogeneous dephasing time $T_2^*$, the relaxation time $T_1$, and the $\pi/2$ pulse duration.

\begin{figure*}[t]
    \centering
    \includegraphics[width=1.0\linewidth]{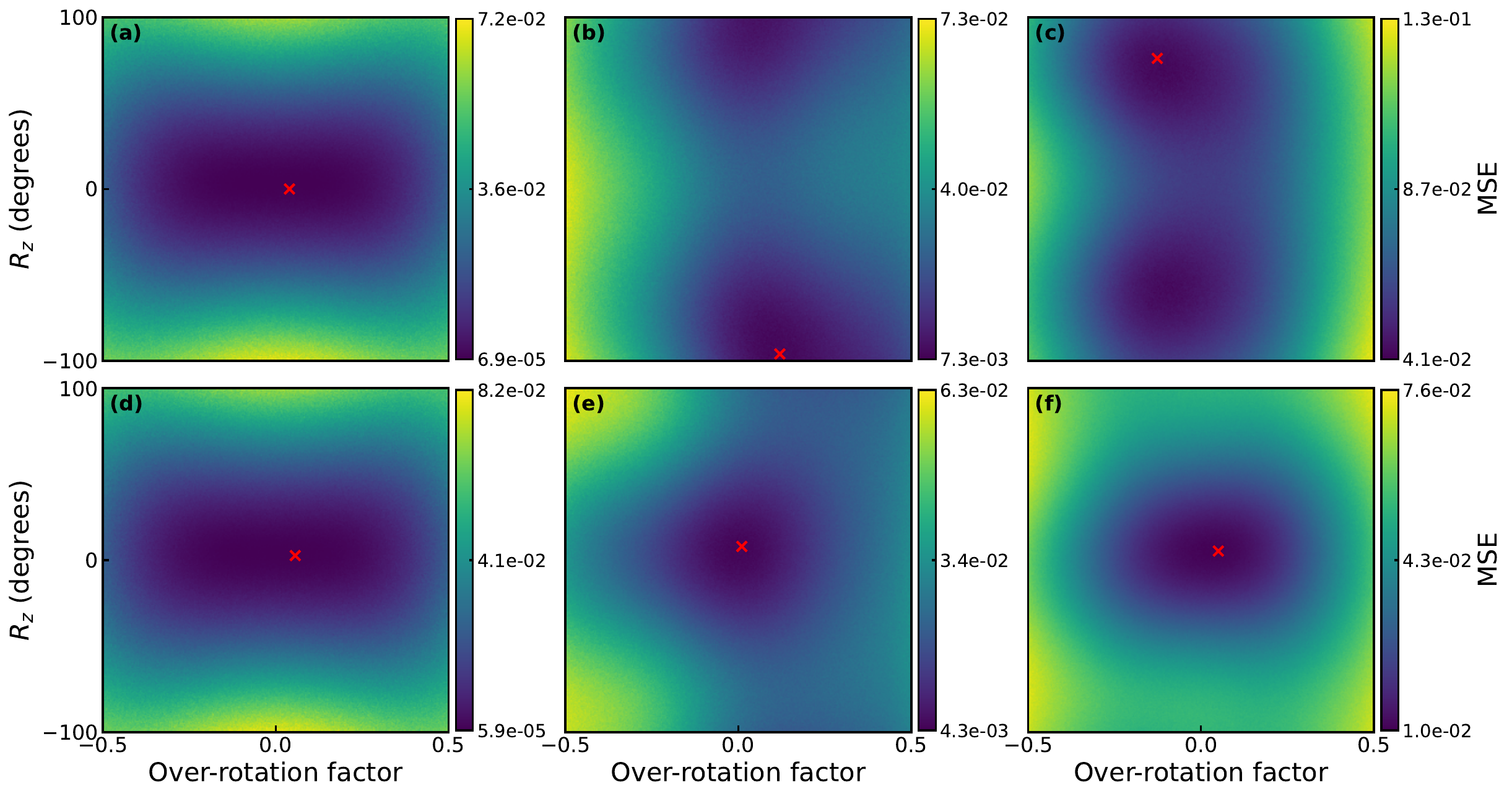}
    \caption{Simulated deviation from ideal evolution (when a standard $\pi/2$ pulse duration of 54~ns is applied) as a function of an over-rotation factor and a phase offset for structured sequences of length 2–4, shown for standard (a–c) and composite (d–f) $\pi/2$ pulses. The over-rotation and phase-offset parameters are swept over a $301 \times 301$ grid, and the color maps represent the mean-squared error (MSE) with respect to the ideal simulated sequence evolution. The points appearing as red crosses indicate the minimum MSE within each parameter landscape. The simulations isolate coherent control errors arising from pulse-duration miscalibration and rotation-axis offsets, highlighting their cumulative effect under repeated application in structured sequences.}
    \label{fig:sim_overrot_phase}
\end{figure*}

This framework allows for the inclusion of different noise models. In particular, to isolate the role of coherent control errors, we introduce controlled perturbations to the ideal pulse operations. Specifically, we consider:
(i) over-rotation errors, implemented as a fractional deviation in pulse duration, and  
(ii) phase errors, implemented as a rotation-axis offset.
These parameters are systematically swept over a grid of $301 \times 301$ points, with phase offsets ranging from $-100^\circ$ to $100^\circ$ and over-rotation factors spanning $-0.5$ to $+0.5$. For each parameter set, we compute the MSE with respect to the ideal sequence evolution, and identify the minimum-error regions shown in Fig.~\ref{fig:sim_overrot_phase}.

The simulations show that coherent errors in pulse duration and phase lead to significant deviations from ideal behavior when pulses are applied sequentially. Compared with standard pulses, composite pulses exhibit lower MSE as the sequence length increases, consistent with the improved suppression of coherent errors observed experimentally in Sec.~\ref{sec:characterization}. However, the simulations also show that, as the sequence length increases, the region of low MSE becomes more localized for composite pulses. This indicates that their improved performance depends on more specific error-compensation conditions rather than being uniformly robust across the parameter space.

While these simulations capture the dominant coherent error mechanisms, quantitative discrepancies remain compared with experiment. These discrepancies arise from additional effects not included in the model, such as ensemble inhomogeneity, control-chain distortions, pulse-shape imperfections, and stochastic noise processes, which are difficult to characterize and incorporate quantitatively into the simulations. As a result, the simulations provide qualitative insight into how coherent errors propagate through multi-pulse sequences rather than a complete quantitative description of the experimental system.

\section{Randomized Benchmarking limitations}
\label{appendix:rb_limitations}

\begin{figure}[t]
    \centering
    \includegraphics[width=0.8\linewidth]{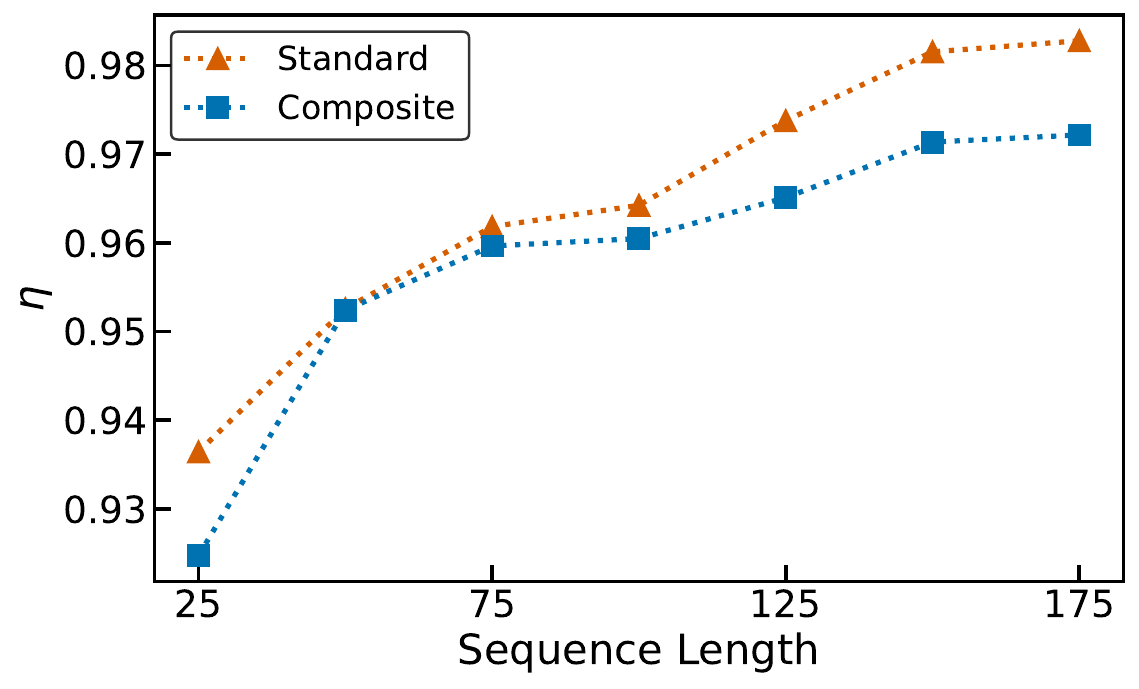}
    \caption{Extracted randomized benchmarking decay parameter $\eta$ as a function of the maximum sequence length included in the fit, obtained from the same dataset shown in Fig.~\ref{fig:RB_normal_composite} for sequence lengths ($n = [25,50,75,100,125,150,175]$).}
    \label{fig:eta_vs_length}
\end{figure}

To examine the dependence of the extracted RB parameter $\eta$ on the fitting range, the fitting procedure described by Eq.~\eqref{eq:RB_model} was repeated while increasing the maximum sequence length included in the fit for the same experimental data set presented in Fig.~\ref{fig:RB_normal_composite}. The resulting values of $\eta$ with sequence lengths are shown in 
Fig.~\ref{fig:eta_vs_length}. A gradual increase in the values of the extracted $\eta$ is observed as longer sequence lengths are included for both pulse implementations. This is counterintuitive given that for sequence lengths $\ge 25$, we are going beyond the $T_2$ time. This indicates that a single exponential process over the complete sequence-length range does not fully describe the RB dynamics in this regime. 

Such behavior is consistent with known limitations of randomized benchmarking in the presence of temporally correlated noise \cite{Fogarty2015, Ng2021, Clerk2025, Wallman2018, Ball2016}, finite-duration gate operations\cite{Feng2016}, and ensemble inhomogeneity \cite{Ryan2009}, where the assumptions underlying ideal Markovian RB are only approximately satisfied. Additionally, errors due to MW power drift and ensemble inhomogeneity as a result of operating with fast pulses are not captured by $\eta$. This motivates the need for RB-adjacent protocols, which can separate control errors from incoherence error channels as well as experimental artifacts.


\section*{Acknowledgments}

The authors thank Lilian Childress and Romain Ruhlmann for valuable discussions, and Romain Ruhlmann, Vincent Halde, and David Roy-Guay from SBQuantum for support and for the use of the Quantum Demonstrator.

\section*{Competing Interests}
A. C-D. has a financial interest in Keysight Technology Inc. and the use of True-Q
software. The other authors declare no conflicting interests.

\bibliographystyle{ieeetr}
\bibliography{ref}

\begin{thebibliography}{10}

\bibitem{Degen2017}
C.~L. Degen, F.~Reinhard, and P.~Cappellaro, ``Quantum sensing,'' {\em Rev. Mod. Phys.}, vol.~89, p.~035002, Jul 2017.

\bibitem{Taylor2008}
J.~M. Taylor, P.~Cappellaro, L.~Childress, L.~Jiang, D.~Budker, P.~R. Hemmer, A.~Yacoby, R.~Walsworth, and M.~D. Lukin, ``High-sensitivity diamond magnetometer with nanoscale resolution,'' {\em Nature Physics}, vol.~4, p.~810–816, Sept. 2008.

\bibitem{Barry2020}
J.~F. Barry, J.~M. Schloss, E.~Bauch, M.~J. Turner, C.~A. Hart, L.~M. Pham, and R.~L. Walsworth, ``Sensitivity optimization for nv-diamond magnetometry,'' {\em Rev. Mod. Phys.}, vol.~92, p.~015004, Mar 2020.

\bibitem{Levchenko2015}
A.~O. Levchenko, V.~V. Vasil'ev, S.~A. Zibrov, A.~S. Zibrov, A.~V. Sivak, and I.~V. Fedotov, ``Inhomogeneous broadening of optically detected magnetic resonance of the ensembles of nitrogen-vacancy centers in diamond by interstitial carbon atoms,'' {\em Applied Physics Letters}, vol.~106, no.~10, p.~102402, 2015.

\bibitem{Blinder2024}
R.~Blinder, Y.~Mindarava, T.~H. Tran, {\em et~al.}, ``Reducing inhomogeneous broadening of spin and optical transitions of nitrogen-vacancy centers in high-pressure, high-temperature diamond,'' {\em Communications Materials}, vol.~5, p.~224, 2024.

\bibitem{Wang2012}
Z.-H. Wang, G.~de~Lange, D.~Rist\`e, R.~Hanson, and V.~V. Dobrovitski, ``Comparison of dynamical decoupling protocols for a nitrogen-vacancy center in diamond,'' {\em Phys. Rev. B}, vol.~85, p.~155204, Apr 2012.

\bibitem{Cory2010}
C.~A. Ryan, J.~S. Hodges, and D.~G. Cory, ``Robust decoupling techniques to extend quantum coherence in diamond,'' {\em Phys. Rev. Lett.}, vol.~105, p.~200402, Nov 2010.

\bibitem{Steiner2010}
M.~Steiner, P.~Neumann, J.~Beck, F.~Jelezko, and J.~Wrachtrup, ``Universal enhancement of the optical readout fidelity of single electron spins at nitrogen-vacancy centers in diamond,'' {\em Phys. Rev. B}, vol.~81, p.~035205, Jan 2010.

\bibitem{Felton2009}
S.~Felton, A.~M. Edmonds, M.~E. Newton, P.~M. Martineau, D.~Fisher, D.~J. Twitchen, and J.~M. Baker, ``Hyperfine interaction in the ground state of the negatively charged nitrogen vacancy center in diamond,'' {\em Phys. Rev. B}, vol.~79, p.~075203, Feb 2009.

\bibitem{Anderson2022}
A.~F.~L. Poulsen, J.~D. Clement, J.~L. Webb, R.~H. Jensen, L.~Troise, K.~Berg-S\o{}rensen, A.~Huck, and U.~L. Andersen, ``Optimal control of a nitrogen-vacancy spin ensemble in diamond for sensing in the pulsed domain,'' {\em Phys. Rev. B}, vol.~106, p.~014202, Jul 2022.

\bibitem{Casanova2025}
I.~n. Iriarte-Zendoia, C.~Munuera-Javaloy, and J.~Casanova, ``Robust microwave cavity control for nv ensemble manipulation,'' {\em Phys. Rev. Res.}, vol.~7, p.~013315, Mar 2025.

\bibitem{McKay2017}
D.~C. McKay, C.~J. Wood, S.~Sheldon, J.~M. Chow, and J.~M. Gambetta, ``Efficient {$Z$} gates for quantum computing,'' {\em Physical Review A}, vol.~96, Aug. 2017.

\bibitem{Barenco1995}
A.~Barenco, C.~H. Bennett, R.~Cleve, D.~P. DiVincenzo, N.~Margolus, P.~Shor, T.~Sleator, J.~A. Smolin, and H.~Weinfurter, ``Elementary gates for quantum computation,'' {\em Phys. Rev. A}, vol.~52, pp.~3457--3467, Nov 1995.

\bibitem{Levitt1986}
M.~H. Levitt, ``Composite pulses,'' {\em Progress in Nuclear Magnetic Resonance Spectroscopy}, vol.~18, no.~2, pp.~61--122, 1986.

\bibitem{Vandersypen2005}
L.~M.~K. Vandersypen and I.~L. Chuang, ``Nmr techniques for quantum control and computation,'' {\em Rev. Mod. Phys.}, vol.~76, pp.~1037--1069, Jan 2005.

\bibitem{Aiello2013}
C.~Aiello, M.~Hirose, and P.~Cappellaro, ``Composite-pulse magnetometry with a solid-state quantum sensor,'' {\em Nature Communications}, vol.~4, p.~1419, 2013.

\bibitem{Dong2021}
Y.~Dong, S.-C. Zhang, H.-B. Lin, X.-D. Chen, W.~Zhu, G.-Z. Wang, G.-C. Guo, and F.-W. Sun, ``Quantifying the performance of multipulse quantum sensing,'' {\em Phys. Rev. B}, vol.~103, p.~104104, Mar 2021.

\bibitem{Dong2022}
Y.~Dong, X.-D. Gao, C.~Yu, Z.-H. Feng, H.-B. Lin, X.-D. Chen, W.~Zhu, and F.-W. Sun, ``Broadband composite pulse for quantum sensing with a solid-state spin in diamond,'' {\em Applied Physics Letters}, vol.~120, no.~19, p.~194001, 2022.

\bibitem{Dong2021_2}
Y.~Dong, J.~Y. Xu, S.~C. Zhang, Y.~Zheng, X.~D. Chen, W.~Zhu, and F.~W. Sun, ``Composite-pulse enhanced room-temperature diamond magnetometry,'' {\em Functional Diamond}, vol.~1, no.~1, pp.~125--134, 2021.

\bibitem{Chakravarty2026}
A.~Chakravarty, R.~Ruhlmann, V.~Halde, D.~Roy-Guay, M.~Pioro-Ladriere, L.~Childress, and Y.~Berube-Lauziere, ``Triple-tone microwave control for sensitivity optimization in compact ensemble nitrogen-vacancy magnetometers,'' {\em Journal of the Optical Society of America B}, vol.~43, pp.~289--299, 2026.

\bibitem{Doherty2013}
M.~W. Doherty, N.~B. Manson, P.~Delaney, F.~Jelezko, J.~Wrachtrup, and L.~C. Hollenberg, ``The nitrogen-vacancy colour centre in diamond,'' {\em Physics Reports}, vol.~528, p.~1–45, July 2013.

\bibitem{Chen2023}
J.~Chen, D.~Ding, C.~Huang, and Q.~Ye, ``Compiling arbitrary single-qubit gates via the phase shifts of microwave pulses,'' {\em Phys. Rev. Res.}, vol.~5, p.~L022031, May 2023.

\bibitem{Lidar2025}
A.~Vezvaee, V.~Tripathi, D.~Kowsari, E.~Levenson-Falk, and D.~A. Lidar, ``Virtual-$z$ gates and symmetric gate compilation,'' {\em PRX Quantum}, vol.~6, p.~020348, Jun 2025.

\bibitem{Wimperis1994}
S.~Wimperis, ``Broadband, narrowband, and passband composite pulses for use in advanced nmr experiments,'' {\em Journal of Magnetic Resonance, Series A}, vol.~109, no.~2, pp.~221--231, 1994.

\bibitem{Cummins2000}
H.~K. Cummins and J.~A. Jones, ``Use of composite rotations to correct systematic errors in nmr quantum computation,'' {\em New Journal of Physics}, vol.~2, p.~006, Mar. 2000.

\bibitem{Zhang2026}
J.~Zhang, C.~K. Cheung, M.~K{\"u}bler, {\em et~al.}, ``Unraveling quantum dephasing of nitrogen-vacancy center ensembles in diamond,'' {\em npj Quantum Materials}, vol.~11, p.~27, 2026.

\bibitem{Knill2008}
E.~Knill, D.~Leibfried, R.~Reichle, J.~Britton, R.~B. Blakestad, J.~D. Jost, C.~Langer, R.~Ozeri, S.~Seidelin, and D.~J. Wineland, ``Randomized benchmarking of quantum gates,'' {\em Phys. Rev. A}, vol.~77, p.~012307, Jan 2008.

\bibitem{Feng2016}
G.~Feng, J.~J. Wallman, B.~Buonacorsi, F.~H. Cho, D.~K. Park, T.~Xin, D.~Lu, J.~Baugh, and R.~Laflamme, ``Estimating the coherence of noise in quantum control of a solid-state qubit,'' {\em Physical Review Letters}, vol.~117, Dec. 2016.

\bibitem{Magesan2012}
E.~Magesan, J.~M. Gambetta, and J.~Emerson, ``Characterizing quantum gates via randomized benchmarking,'' {\em Physical Review A}, vol.~85, Apr. 2012.

\bibitem{Wallman2018}
J.~J. Wallman, ``Randomized benchmarking with gate-dependent noise,'' {\em Quantum}, vol.~2, p.~47, Jan. 2018.

\bibitem{Proctor2017}
T.~Proctor, K.~Rudinger, K.~Young, M.~Sarovar, and R.~Blume-Kohout, ``What randomized benchmarking actually measures,'' {\em Physical Review Letters}, vol.~119, Sep 2017.

\bibitem{deLange2010}
G.~de~Lange, Z.~H. Wang, D.~Ristè, V.~V. Dobrovitski, and R.~Hanson, ``Universal dynamical decoupling of a single solid-state spin from a spin bath,'' {\em Science}, vol.~330, p.~60–63, Oct. 2010.

\bibitem{Pham2012}
L.~M. Pham, N.~Bar-Gill, C.~Belthangady, D.~Le~Sage, P.~Cappellaro, M.~D. Lukin, A.~Yacoby, and R.~L. Walsworth, ``Enhanced solid-state multispin metrology using dynamical decoupling,'' {\em Physical Review B}, vol.~86, July 2012.

\bibitem{Erhard2019}
A.~Erhard, J.~J. Wallman, L.~Postler, M.~Meth, R.~Stricker, E.~A. Martinez, P.~Schindler, T.~Monz, R.~Blatt, and N.~Friis, ``Characterizing large-scale quantum computers via cycle benchmarking,'' {\em Nature Communications}, vol.~10, p.~5347, 2019.

\bibitem{Wallman2015}
J.~Wallman, C.~Granade, R.~Harper, and S.~T. Flammia, ``Estimating the coherence of noise,'' {\em New Journal of Physics}, vol.~17, p.~113020, Nov. 2015.

\bibitem{Rembold2020}
P.~Rembold, N.~Oshnik, M.~M. Müller, S.~Montangero, T.~Calarco, and E.~Neu, ``Introduction to quantum optimal control for quantum sensing with nitrogen-vacancy centers in diamond,'' {\em AVS Quantum Science}, vol.~2, Jun 2020.

\bibitem{Beale2020}
S.~J. Beale, ``{True-Q},'' June 2020.

\bibitem{Fogarty2015}
M.~A. Fogarty, M.~Veldhorst, R.~Harper, C.~H. Yang, S.~D. Bartlett, S.~T. Flammia, and A.~S. Dzurak, ``Nonexponential fidelity decay in randomized benchmarking with low-frequency noise,'' {\em Phys. Rev. A}, vol.~92, p.~022326, Aug 2015.

\bibitem{Ng2021}
J.~Qi and H.~K. Ng, ``Randomized benchmarking in the presence of time-correlated dephasing noise,'' {\em Phys. Rev. A}, vol.~103, p.~022607, Feb 2021.

\bibitem{Clerk2025}
A.~Brillant, P.~Groszkowski, A.~Seif, J.~Koch, and A.~A. Clerk, ``Randomized benchmarking with non-markovian noise and realistic finite-time gates,'' {\em Phys. Rev. Lett.}, vol.~135, p.~070601, Aug 2025.

\bibitem{Ball2016}
H.~Ball, T.~M. Stace, S.~T. Flammia, and M.~J. Biercuk, ``Effect of noise correlations on randomized benchmarking,'' {\em Physical Review A}, vol.~93, Feb. 2016.

\bibitem{Ryan2009}
C.~A. Ryan, M.~Laforest, and R.~Laflamme, ``Randomized benchmarking of single- and multi-qubit control in liquid-state nmr quantum information processing,'' {\em New Journal of Physics}, vol.~11, p.~013034, Jan. 2009.

\end{thebibliography}

\end{document}